\documentclass[nofootinbib,a4paper,aps,pra,showpacs,preprintnumbers,twocolumn]{revtex4-2}

\usepackage[T2A]{fontenc}
\usepackage{amssymb}
\usepackage{amsmath}
\usepackage{mathtools}
\usepackage{braket}
\usepackage{graphicx}
\usepackage{epstopdf}
\usepackage[english]{babel}
\usepackage{url}
\usepackage{listings}
\usepackage{gensymb}
\usepackage{xcolor}
\usepackage{ulem}
\usepackage{enumitem}

\newlist{steps}{enumerate}{1}
\setlist[steps, 1]{label = Step \arabic*:}

\begin{document}

\title{Heralded gate search with genetic algorithms for quantum computation}
\author{A. Chernikov$^1$, S.S. Sysoev$^2$, E.A. Vashukevich$^1$, T.Yu. Golubeva$^1$}
\affiliation{\center{${}^1$ Saint-Petersburg\;State\;University,}\newline ${}^2$ Leonhard\;Euler\;International\;Mathematical\;Institute}
\begin{abstract}
In this paper we present genetic algorithms based search technique for the linear optics schemes, performing two-qubit quantum gates. We successfully applied this technique for finding heralded two-qubit gates and obtained the new schemes with performance parameters equal to the best currently known. The new simple metrics is introduced which enables comparison of schemes with different heralding mechanisms. The scheme performance degradation is discussed for the cases when detectors in the heralding part of the scheme are not photon-number-resolving. We propose a procedure for overcoming this drawback which allows us to restore the reliable heralding signal even with not-photon-number-resolving detectors.   
 \end{abstract}

\maketitle

\hyphenation{}
	
\section{Introduction}

For more than 20 years photonic quantum-computational networks development remains a challenging task for the scientific community \cite{Slus}. Photons as a platform for quantum information attract attention for various reasons: they are inexpensive, easily manipulated, reliably detected, transmitted and they can even be stored in quantum memory. Photons are our best choice for the task of avoiding decoherence, since they almost never interact with each other. The latter though appears to be the source of the main obstacle when it comes to computations, since this interaction is necessary for controlled quantum gates.  

The use of linear optics schemes with heralding is currently believed to be the most beneficial approach for the development of photonic quantum gates \cite{Kok}. Such schemes are based on the principle of probabilistic gate actuation with a probability much lower than 100\%. Performing the transformation on quantum systems of a larger dimension, linear optical circuits involve the detection of a certain pattern at the output of the circuit (heralding signal) indicating that the required operation is performed on unmeasured outputs. All detection outcomes in which this pattern is not implemented are considered unsatisfactory, which leads to a significant decrease in the gate actuation probability. Low actuation probabilities prevent effective heralded schemes scaling. To improve this situation, it is proposed to use several identical schemes in place of one with the selection of correct outputs for further computation \cite{Holger}. However, this approach is not only resource-intensive (it involves a significant increase in the number of optical elements), butbut it also requires signal delaying and use of fast switches. All these elements become a source of channel losses and the additional noise.

The first linear optics entangling two-qubit gate was introduced in 2001 by Knill, Laflamme and Milburn \cite{KLM}, among with the KLM protocol, named after the names of its authors. This $CZ$ gate has an actuation probability of 1/16 and heralding mechanism implemented with 4 ancillary spatial modes and 2 ancillary photons. This result inspired a great optimism and interest in the linear optics gate implementations, which in less than a year resulted in better implementation \cite{Knill2002} of $CZ$ with an actuation probability of 2/27 and a much simpler scheme architecture. Later in \cite{CNOT2003} another authors achieved even better actuation probability of 1/9, though with a less reliable heralding event with post-selection procedure.

The later progress in the area produced more gate architectures \cite{Langford, Okamoto, Gazzano}. It was also proposed to use entangled ancillary photons \cite{Pittman, Zeuner18, Zhao, Gasparoni} and Bell states measurements \cite{Li2021, Gao} to achieve higher probabilities. Schemes using auxiliary entangled photon states provide extremely high probabilities (up to 1/4 \cite{Gasparoni}). However, if we include the input states generation into scheme analysis, the advantage of such schemes does not seem obvious due to the low number of operations per unit of time due to the necessity of generation of a multi-photon state.

In this paper we exclude schemes with pre-entangled photons from consideration and concentrate on two types of  heralding event: detection pattern in ancillary modes (like in \cite{KLM, Knill2002}) and post-selection with photons detection in signal (non-ancillary) modes (like in \cite{CNOT2003}). 

Finding an effective optical scheme for a conditional gate has proven to be a complicated task, and the modern approach for handling complicated tasks is feeding them to a computer. With several natural restrictions the task can be formulated as a global optimization problem, for which plenty of heuristics are available at the moment. We chose genetic algorithms, introduced in \cite{GA1, GA2} and fruitfully employed for a diverse set of tasks ever since (see, for example \cite{GA_REC_1, GA_REC_2, GA_REC_3, GA_REC_4}).  

Here we present the results of the genetic algorithm application to the problem of finding the conditional gate scheme with best fidelity and probability in the KLM-protocol. 

We consider two well-known schemes that implement entangling gates in order to identify their features and discuss the various probabilities that characterize the gates. This opens up two different search strategies that we are implementing in this paper. In addition, we analyzed the stability of the scheme to detection errors. We limited the outcomes to the act of distinguishing of just two events on the detectors in the heralding channels: the absence of a click corresponds to the vacuum state of the field in the channel, the presence of a click indicates a Fock state with one photon or more. This limitation corresponds to the modern capabilities of measuring equipment.

The algorithm we have built allows us to consider the measurement possibilities in the optimal scheme search.

The paper is structured as follows: in the section II we analyze some well-known schemes in the KLM-protocol. We describe the heralding procedure in terms of conditional probabilities, which reveals parameters for further optimization. We also discuss the change in heralding characteristics when the detector is a photon-number-resolving (PNR). The section III is fully dedicated to the genetic algorithms and their implementation for the task of optical schemes search. We describe two series of experiments: optimizing the total actuation probability with no regard to the quality of heralding signal, and optimizing the actuation probability with absolutely reliable heralding. Both series of experiments produced schemes which we analyze in the subsections IIID and IIIE. In conclusion we present our interpretation of these results.

All optical schemes discussed in this paper were analyzed and tested in LOQC TECH web-application \cite{LOQC}.   

\noindent
\section{Probability analysis in KLM-Protocol}
\subsection{Qubits encoding and the state space}
Building a universal quantum computer means fulfilling the set of conditions called DiVincenzo criteria \cite{DiVin}. Photonic systems painlessly satisfy most of them leaving the only one for consideration -- the implementation of universal set of quantum gates \cite{Barenko}. For qubits this universal set consists of three one-qubit rotations on the Bloch sphere and one two-qubit entangling gate. The choice of photons as a platform for computations makes single qubit operations easy with wave-plates and polarising beam splitters. Two-qubits entanglement on the contrary seems hard if even possible, since photons don't interact with each other under normal conditions. This interaction can in principle be achieved through non-linear optical medium \cite{Wang}, but it is too weak to be considered as a gate implementation resource. The most promising approach for the task of two qubit gates implementation on photons is KLM-protocol \cite{KLM}, where photon interference occur on linear optical elements and the gate actuation has essentially probabilistic  nature. 
 
In KLM-protocol each qubit is encoded as a photon location in two spatial modes. $\ket{0}$ corresponds to the photon being in one mode, $\ket{1}$ -- in another. This encoding can be easily obtained from polarization encoding, where horizontal polarization of a photon encodes $\ket{0}$ and vertical -- $\ket{1}$.

The spatial modes which encode qubits (the signal modes) are supplemented by ancillary modes, possibly carrying ancillary photons. The particular pattern of photons detection on the output of ancillary modes represents heralding signal of the gate actuation. The number of ancillary modes and ancillary photons depends on the gate implementation. They expand the state space of the system, and the detection of ancillary photons on the output reduces it back to that represented by signal modes.

The photons in the signal modes can violate the qubits encoding by grouping in one mode or leaving to ancillary modes. Despite the fact that the presence of two photons in the signal mode at the output indicates that the gate is actuated incorrectly and takes the quantum state out of the two-qubit space, the phenomenon of photon bunching itself is an integral part of quantum interference and is always present in the schemes of entangling operations. 

The evolution of quantum state is performed by linear optical elements -- beam splitters and phase-shifters. A phase-shifter $PS_{\phi}$ acts on one mode by changing its phase: $PS_{\phi}\ket{x} = e^{i\phi}\ket{x}$. A beam splitter $BS_{\theta,\phi}$ acts on two modes, performing the following unitary transform:

\begin{equation}
BS_{\theta, \phi}=\begin{pmatrix}
		\cos{\theta} & -e^{i\phi}\sin{\theta} \\
		e^{-i\phi}\sin{\theta} & \cos{\theta}
	\end{pmatrix}.
\end{equation}

 Angle $\theta$ thereby affects the beam splitter transparency with $\cos{\theta}$ being the reflection amplitude coefficient  and $\sin{\theta}$ -- the transmission amplitude coefficient. Angle $\phi$ represents the relative phase shift between two outcomes.

Beam splitter acts on two different modes, so it is our main resource for entanglement. Gate CZ from \cite{KLM} (we will refer to this gate as CZ(1/16) in future analysis and gates comparisons), for example, uses $BS_{45\degree,0}$ to collect photons from two different modes in one mode (the effect of photon bunching primarily discovered in \cite{HOM}). In this case, as mentioned earlier, we go beyond the two-qubit logic, since in many schemes of entangling transformations, the dimension of the Hilbert space used is greater than the dimension of the two-qubit space. To eliminate this feature, we must be able to distinguish and discard those output states that are not the desired result of the gate action.

\subsection{Heralding event based on photons detection}
\begin{figure}[t]
		\centering
	\includegraphics[width=3.5in]{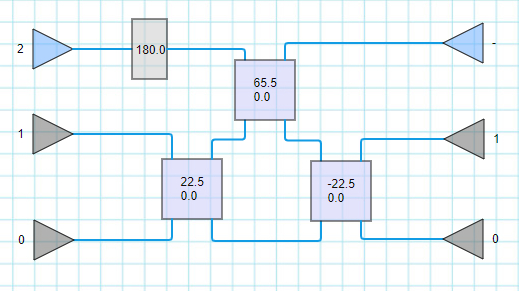}
	\caption{The NSx Gate. Our notation for the schemes is as follows: blue triangles denote inputs and outputs of the signal modes, grey triangles represent inputs and outputs of ancillary modes. Photons paths go from left to the right. Numbers near the inputs denote the number of input photons in the corresponding mode. Numbers near the ancillary outputs specify the number of photons to be detected for the correct heralding signal (the heralding pattern). Grey rectangles represent phase-shifters with angle $\phi$ specified inside the rectangle. Purple squares stand for beam splitters with $\theta$ in the first row and $\phi$ in the second.}
	\label{NS_x}
\end{figure}

Let's consider the heralding mechanism with example of the $NS_x$ gate from \cite{KLM} (Fig. \ref{NS_x}). It has one signal mode (the first one, with blue triangle denoting its input) and two ancillary modes (2 and 3, with grey triangle inputs). The state in the first mode is only transformed when it has 2 photons, in which case it changes its phase by $\pi$. The correct gate actuation coincides with the particular detection pattern on ancillary modes.

\begin{figure}[t]
    \centering
    \includegraphics[width=3.5in]{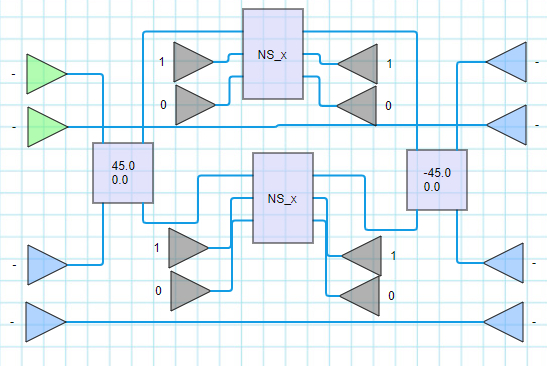}
    \caption{Gate CZ(1/16), based on 2 NSx gates. Green inputs denote control qubit modes.}
    \label{figNSXCX}
\end{figure}

This detection pattern indicates 1 photon in the mode number 2 and 0 photons in the mode number 3. When this heralding signal is obtained one can be sure about the correct gate actuation, which for this particular scheme happens with probability of $\frac{1}{4}$. Gate CZ(1/16) uses two $NS_x$ gates (Fig. \ref{figNSXCX}).

The heralding pattern of CZ(1/16) is formed of heralding patterns of both $NS_x$, thus giving the probability of correct heralding 1/16. We arrive here to an important point -- in general there's a difference between probability of correct heralding signal  and the correct gate actuation. Let's introduce the following notation: $P$ -- the probability of correct actuation, $P_a$ -- the probability of correct heralding signal, $P_b$ -- the conditional probability of correct actuation when the correct heralding event happened. Then  
$$
P = P_a P_b
$$
If photon number resolving detectors are used for heralding in CZ(1/16), then for this scheme $P = P_a = 1/16$, with $P_b = 1$.

\subsection{Heralding event based on coincidence registration}\label{2C}

\begin{figure}[t]
    \centering
    \includegraphics[width=3.5in]{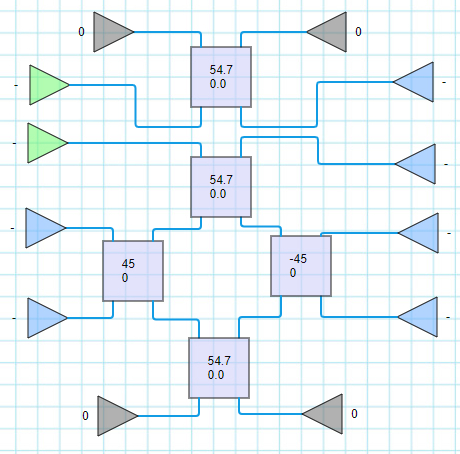}
    \caption{Scheme of CX(1/9).}
    \label{figCX}
\end{figure}

Let's consider the scheme of gate $CX$ introduced in \cite{CNOT2003}. This gate has better actuation probability -- 1/9 (Fig. \ref{figCX}, hereafter we'll refer it as CX(1/9)).

\begin{figure}[t]
    \centering
    \includegraphics[width=3.5in]{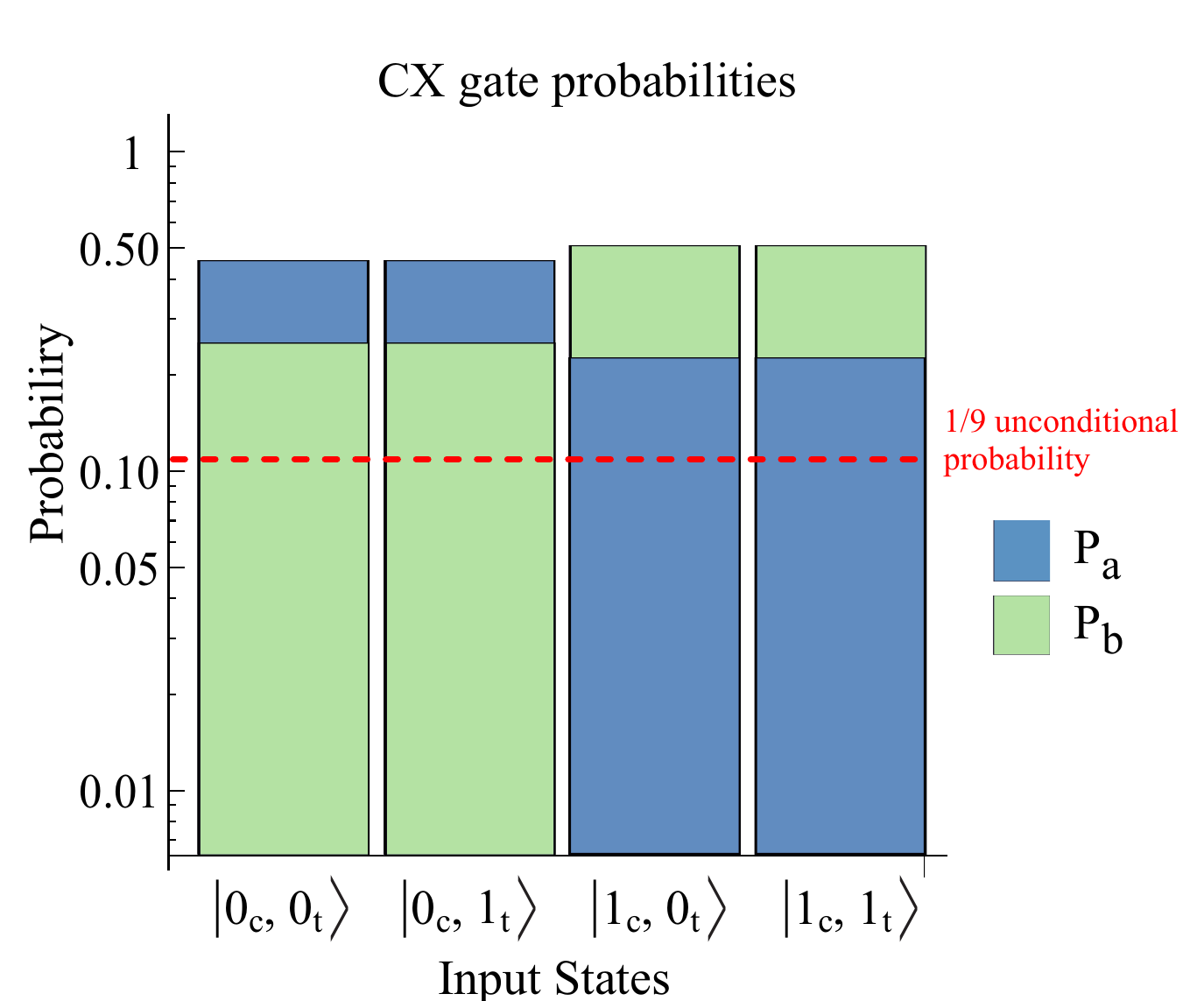}
    \caption{Probabilities $P_a$ and $P_b$ for CX(1/9). Heralding pattern is absence of photons in ancillary outputs.}
    \label{figProbCX19}
\end{figure}

However, this actuation probability doesn't fully rely on heralding, since the correct heralding signal (zero photons in the auxiliary modes) for this scheme does not accompanied by 100\% correct actuation, and $P_b<1$ (see Fig. \ref{figProbCX19}). 

Thus for the input states $\ket{0_c, 0_t}$ or $\ket{0_c, 1_t}$ (subscript "c" denotes control qubit, "t" -- target) the conditional probability of correct actuation when the correct heralding event happened $P_b=0.25$, and for inputs $\ket{1_c, 0_t}$ or $\ket{1_c, 1_t}$, $P_b=0.5$. Unlike in the case of CZ(1/16), the heralding pattern alone is unreliable for this gate.

The authors  \cite{CNOT2003} proposed another heralding mechanism –- that is to distinguish the success of the operation by the presence of coincidences between the acts of detecting single photons in spatial modes encoding the
states of qubits. This leads to a significant inconvenience when performing computational operations – one cannot be sure whether the gate has actuated correctly until the end of the entire calculation (which may consist of many gates) . In addition, the measurement of coincidences tightens the requirements for signal synchronization. In gates which rely on photon counting, we have to ensure synchronizationonly for the elements with 2-photon interference, but in circuits with a coincidence count, signal synchronization should be maintained until the final measurement.

Moreover the schemes with coincidental heralding are difficult to scale. With a cascade of several gates, the presence of coincidences between photocounts in signal modes does not guarantee the correct operation of the entire cascade, since the correct heralding signal may be revealed when one of the gates did not work correctly as well in the case of the correct operation of the entire circuit (we will discuss this problem in more detail in Section III (F).

\subsection{Photons detection}\label{2D}

\begin{figure}[t]
    \centering
    \includegraphics[width=3.5in]{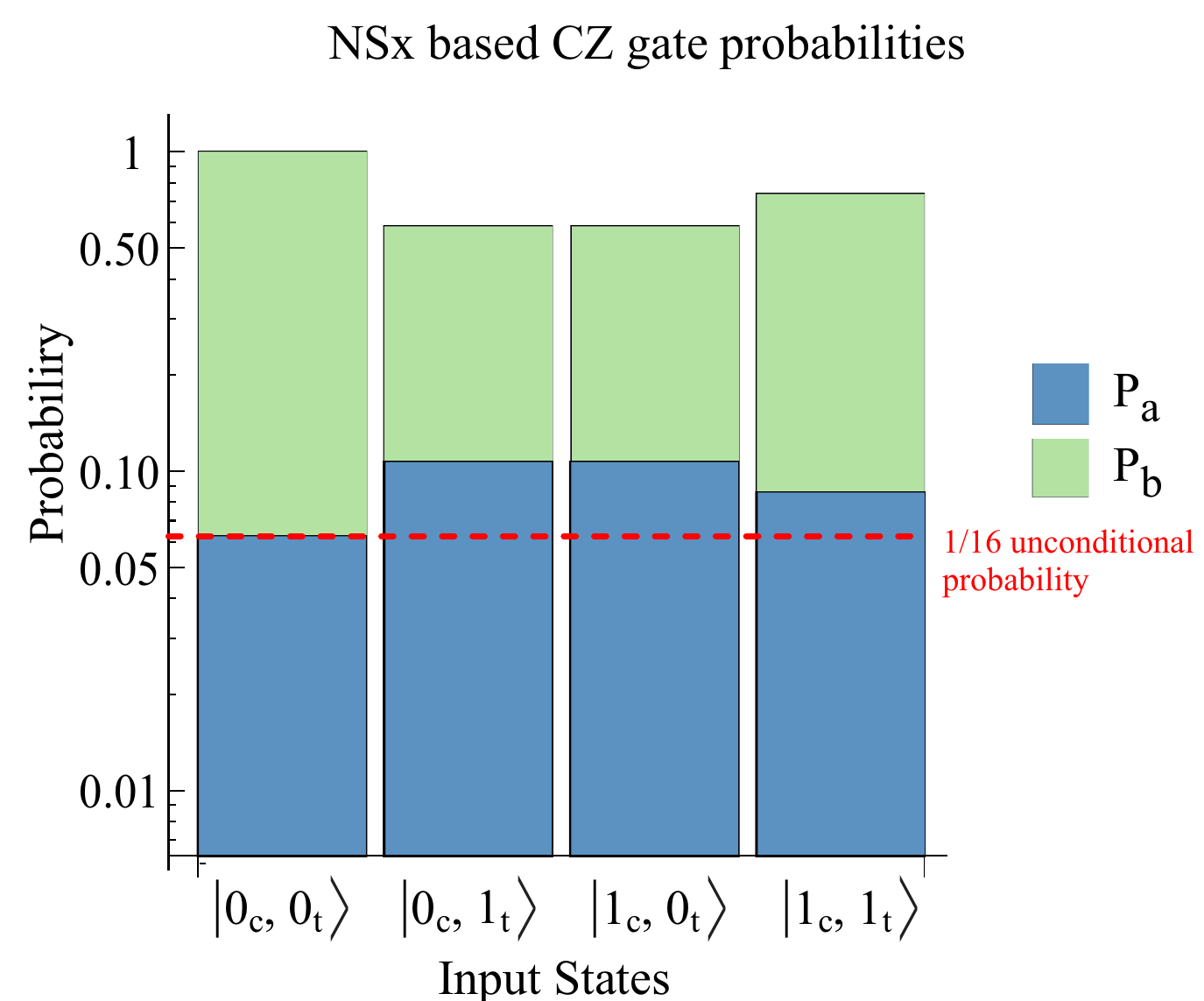}
    \caption{$P_a$ and $P_b$ of CZ(1/16) for the case of non-photon-number-resolving detectors.}
    \label{figProbCZ}
\end{figure}

 Let us noting that photon-number-resolving detectors are not always an undoubtedly implementable resource. It is more realistic to consider possibility of distinguishing just two outcomes -- one (or more) photons detected or none photons detected in one mode. In this case the probability $P$ of CZ(1/16) actuation doesn't change, but its components, $P_a$ and $P_b$ shift and become dependent of the input state (Fig. \ref{figProbCZ}). If we can't discern one photon from two on the output of the ancillary modes, the heralding signal becomes less reliable. 
 
 From figure \ref{figProbCZ} one can conclude that for the input $\ket{0_c, 0_t}$ probabilities $P_a$ and $P_b$ don't change, but for other inputs they do, and increased probability of correct heralding event $P_a$ decreases the conditional probability $P_b$ of correct gate actuation. This depreciates the superiority of CZ(1/16) over CX(1/9) in terms of reliable heralding, possibly making the latter more appealing for the reason of better actuation probability.

 Thus we find ourselves in need of a criteria for gates evaluation based on their actuation probability components. The previous reasoning shows that the total probability of the gate triggering is not a sufficient parameter for optimizing the scheme, The conditional probability $P_b$ is an important parameter for scaling, but as it was shown earlier, it can depend differently from scheme input for the different schemes. Lowest and highest values of $P_b$ on the inputs of computational basis can provide some insight on the scheme heralding mechanism performance, but they become much less useful if the logical basis changes. 

It is easy to show that the mean value of $P_a$ over any orthonormal basis is independent from basis choice. Let's denote it $\bar{P_a}$:
 $$
 \bar{P_a} = \frac{P_a(\ket{00}) + P_a(\ket{01}) + P_a(\ket{10}) + P_a(\ket{11})}{4}
 $$

If the input of the scheme is considered as a random event with uniform distribution then $\bar{P_a}$ becomes the expected value of correct heralding signal. Most quantum algorithms perform Hadamard transform in the start of computation, thus preparing this uniform distribution of all possible inputs in computational basis. In this case the probability of heralding event $P_a$ equals $\bar{P_a}$.  

For the conditional gates considered in this paper the conditional probability $P_b$ is a function of $P_a$:
$$
P_b = P/P_a
$$
Let's introduce the $\bar{P_b}$ -- the expected value of correct gate actuation with correct heralding signal for the case of uniform distribution of inputs:
$$
\bar{P_b} = P/\bar{P_a} = E({P_b})
$$
Based on the above we choose $\bar{P_b}$ to be our criteria for scheme evaluation. For example, for the gate CX(1/9) it can be computed as follows:
$$
\bar{P_a} = \frac{2/3 + 2/3 + 2/9 + 2/9}{4} = 4/9 
$$
$$
\bar{P_b} = \frac{1/9}{4/9} = 1/4
$$
We will use these characteristics to evaluate the proposed transformation schemes further.
\section{Genetic Algorithms for the linear optics quantum gates search}
\subsection{Principle of operation of genetic algorithms}
Genetic (or evolutionary) algorithm introduced in \cite{GA1,GA2} is the heuristic approach for the global optimization problem. It benefits from the idea of natural evolution, which seems very successful in optimization of living species in nature. The simple idea of the algorithm is this:

\begin{steps}
	\item{Initialization. Produce the set of random species -- the elements of the search space (the first generation).}
	\item{Selection. Evaluate each specie in the generation (according to some fitness function) and choose the subset of the best of them.}
	\item{Crossover. Produce the offspring (the next generation) of the selected species by random recombination of their genetic material.}
	\item{Mutations. Apply random changes to the species of the new generation.}
	\item{Go to step 2 with the new generation.}
\end{steps}

\subsection{Tuning of the numerical experiments parameters}

In our experiments the species represent optical schemes for conditional quantum gates. Each such scheme has four signal and arbitrary number of ancillary modes. Four signal modes represent two qubits -- the control and the target one. For each experiment several parameters were chosen:

\begin{itemize}
    \item{Generation size $G$ is the number of species in one generation.}
	\item{Number of parents $G_p$ is the number of the best species in the generation that are chosen to produce offspring.}
	\item{Mutation rate $R$ corresponds to the probability of random changes of each specie during the mutation step.}
	\item{Scheme depth (or complexity) $d$ accounted with the number of beam splitters in each scheme.}
	\item{The number of ancillary modes $N_a$.}
	\item{The number of photons in ancillary modes $n_a$.}
\end{itemize}

\noindent
Each optical scheme is represented by an array of numbers (the specie's genome). This genome consists of several parts each responsible for a particular information about a scheme:

\begin{itemize}
	\item{$\theta$ rotation angles: $d$ numbers, one for each beam splitter in the represented circuit.}
	\item{$\phi$ rotation angles: phase rotations in beam splitters and phase-shifters.}
	\item{Modes numbers: $d$ pairs of numbers specifying the input modes for each beam splitter and up to $2d$ numbers for phase-shifters.}
	\item{Ancillary inputs: $n_a$ mode numbers, one for each ancillary photon input.}
	\item{Ancillary outputs: $n_a$ mode numbers, one for each ancillary photon detection.}
\end{itemize}

\noindent
Rotation angles are specified in grades with precision up to 2 decimal points. 

\noindent
On initialization step we generate $G$ random genomes representing the optical circuits. 

\noindent
Selection step requires a fitness function for species evaluation. We imply this function optimizes two values -- the fidelity of a conditional gate scheme and the probability of the gate actuation. 

In the first set of experiments, described in subsection \ref{3C}, our goal was to find the best actuation probability $P$ among high fidelity schemes. In that part of research we set aside heralding probability $P_a$ and conditional probability $P_b$. Then, in the second set of experiments (subsection \ref{3E}), we concentrated on the search of schemes with good heralding, those having $P_b = 1$. The goal was to find the best probability of schemes which allow scaling. In fact, such a search gives an answer to the question, what is the maximum probability $P$ of an actuation of a two qubit gate that works with a 100\% probability if the correct heralding event occurs.

\noindent
For the crossover step we also chose different approaches for different sets of experiments: 

\begin{itemize}
	\item{Single-point crossover, which selects a random point in genome and generates the child by taking all genetic material before that point from parent 1, and the remainder -- from parent 2.}
	\item{Uniform crossover -- parent is chosen randomly for each gene.}
\end{itemize}

\noindent
Single-point crossover have proven to be the best choice for the first set of experiments, since the information about each optical element was stored locally in scheme's genome. Splitting genome in just one random point caused high probability of passing successful parts of the scheme to the offspring as a whole. 
In the second set of experiments we modified the genome encoding to improve the speed of scheme analysis. The description of each optical element became scattered across the whole scheme thus making uniform crossover the most natural choice.

\noindent
Our code for all experiments is available on GitHub \cite{galopy}. In the first set of experiments we used  PyGAD library \cite{PyGAD} for the genetic algorithms, then we had to develop our own implementation to enable CUDA. Development language -- Python 3.8.  

\noindent
The first set of experiments (on CPU) were run on AMD Ryzen 5 3600 6-core processor with 32Gb memory. GPU was not used.
For the second set (with CUDA and GPU) we used Google Colab with GPU-enabled runtime.

\begin{figure}[t]
	\centering
	\includegraphics[width=3.5in]{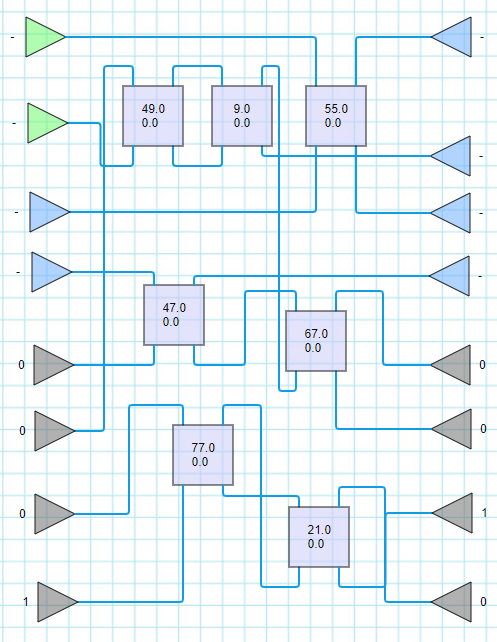}
	\caption{CZ-gate, fidelity F=0.99187..., probability P=0.0635... Obtained after 189 generations, 99 computation hours. } 
	\label{galopy1}
\end{figure}

\begin{figure}[t]	
	\centering
	\includegraphics[width=3.5in]{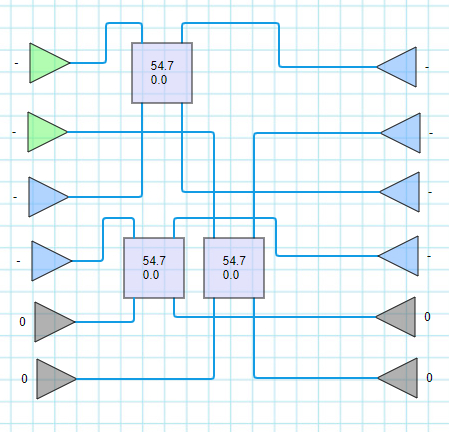}
	\caption{$CZ$-gate, fidelity $F$ = 1, actuation probability $P$ = 1/9. } 
	\label{CZ93}
\end{figure}

\subsection{Experiment set 1: actuation probability optimization}\label{3C}

For the first set of experiments we used generation size $G=5000$, number of mating parents $G_p=800$, mutation rate introduced with assigning random values to the quarter of genes with probability 0.5. We started with scheme complexity $d=8$ and number of ancillary modes $N_a=4$. Numbers $d$ and $N_a$ were chosen to allow the gate CZ(1/16) or similar to be found. 

Single-point crossover were used for the crossover step.

We needed the fitness function $f_1$, used on the selection step, to optimize two parameters -- actuation probability $P$ and fidelity $F$. We used different approaches to combine these values, including multiplying them in different powers and exponentiation, but the best results were achieved with the threshold function, that is to optimize one parameter up to a certain threshold, and then optimize the other.

\begin{lstlisting}
def f_1(P, F):
    if P < P_min:
        return P
    return 1000 * F
\end{lstlisting}

Function $f_1$ defined this way strongly favors fidelity over probability and excludes all low probability schemes ($P < P_{min}$) from consideration by giving them low fitness values. For the schemes with affordable probability $P \geq P_{min}$ better fidelity wins regardless of probability. We set $P_{min} = 0.1$ to be able to find gates like CX(1/9) or better.

The stopping criteria for the algorithm was reaching the fidelity value of 99\%. After that we used LOQC.TECH \cite{LOQC} for the best schemes analysis and tuning. After few experiments we excluded phase-shifters from genome representation and replaced them with beam splitters, since phase-shifters didn't affect the entanglement logic and occupied the significant part of genome, consuming computing resources. 

The first issue we noted was that chosen set of parameters produce significant genetic garbage (that is optical elements which do not affect the scheme performance). One of the examples can be found on Fig. \ref{galopy1}. This looks very natural since evolution doesn't have instruments for eliminating of something that doesn't affect survival.  

The other notable point was that literally all schemes that survive selection have beam splitters on the modes 1 and 3 with $\theta$ close to 50-60 degrees. Removing insignificant beam splitters and attuning angles of the remained ones gave us the scheme with good fitness value just on 3 beam splitters. 

Since very extensive search in the space of schemes with at most 8 optical elements gave us the scheme with just 3 beam splitters, we decided to repeat the experiment with at most 3 beam splitters. Our reasoning goes as follows:

\begin{itemize}
	\item {3 beamsplitters is the minimal number of entangling elements for entangling 2 qubits in 4 spatial modes.}
	\item {More beam splitters produce more wrong paths for photons to go thus reducing the actuation probability of a gate.}
\end{itemize}

We also increased the precision for $\theta$ by 1 decimal digit. After 11 days the algorithm stopped on reaching stopping criteria for fidelity = 0.999 and produced the scheme on Fig. \ref{CZ93}. This scheme entangles the first modes of control and target qubits, while the second modes of both qubits are entangled each with its own ancillary mode. With angles $\theta = \arccos{\frac{1}{\sqrt{3}}}$ it reaches the best possible fidelity $F=1$  with actuation probability $P=1/9$ and implements the CZ transform. 

 We analysed this found gate (hereafter we refer it  -- CZ(1/9)) with the new metric introduced earlier in Sec. IIE   and obtained the same results as for CX(1/9), though the values of $P_a$ are slightly different on the basis inputs:
 
$$
P_a(\ket{00}) = 1/9
$$
$$
P_a(\ket{01}) = 1/3
$$
$$
P_a(\ket{10}) = 1/3
$$
$$
P_a(\ket{11}) = 1
$$
$$
\bar{P_a} = \frac{1 + 1/9 + 1/3 + 1/3}{4} = 4/9 
$$
$$
\bar{P_b} = \frac{1/9}{4/9} = 1/4
$$  

This result gave us certainty about the applicability of the chosen search technique for our task.

\subsection{Experiments with gradient descent}\label{3D}

For the second set of experiments we planned search of schemes with good heralding -- those having $P_b$ equal to 1 or close to it on all legitimate inputs (like CZ(1/16)). However with the settings of the first experiments set we found nothing close to this goal. The search space was too large and the iterations -- too slow. We chose two different approaches to address this problem: to optimize the algorithm implementation and to narrow the search space. The second approach is the subject of this subsection.

To narrow the search space we needed to find out at least something about the search goal. 
Genetic algorithm is an heuristic which allows to address the problem of global optimization, and as the majority of heuristics it is poorly justified. There is a set of well justified methods based on the idea of moving in the search space along the direction of the loss function gradient. These are gradient descent and various pseudo-gradient methods. Unfortunately they are well justified only for the task of local optimization and with restrictions on the loss function. 

The loss function must at least be differentiable for all its variables. This restriction doesn't allow us to use gradient methods for the search of scheme design (the places of optical elements on the scheme), since this design is represented by integers -- mode numbers. The search, however, is possible, when the design is fixed and only the real parameters of the optical elements have to be attuned. The fixed design therefore must allow representation of any possible unitary transform, which means it has to be universal. Finding gate representation in universal design by gradient descent could allow us to form some knowledge about the parameters of the scheme for further search with genetic algorithms.

In this set of experiments we used universal designs introduced by Reck \cite{Reck}, Clements \cite{Clements} and our own, close to Reck design but with more convenient expression for the loss function. The initial scheme size was based on the scheme of CZ(1/16) -- four signal and four ancillary modes with two ancillary photons.   

\begin{figure}[t]	
	\centering
	\includegraphics[width=3.5in]{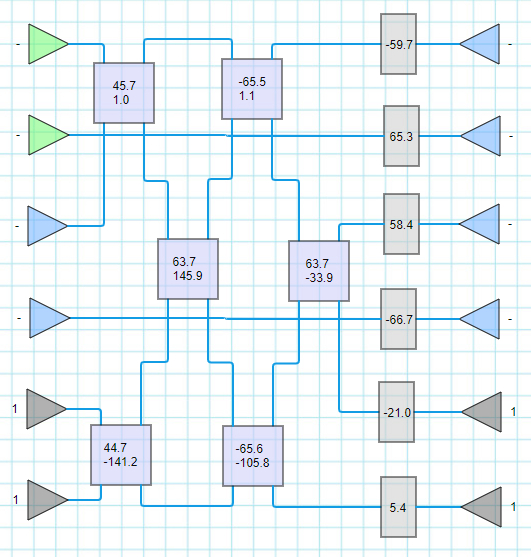}
	\caption{Gradient descent result in Clements design, probability P = 2/27, fidelity F = 1.} 
	\label{CZClements}
\end{figure}

Local optimization methods make the choice of the initial point of iterations very important. Our strategy was to choose this point randomly for each launch, thus the vast majority of experiments ended up in local minima without reaching tolerable values of fidelity. It took some time before we found a scheme with fidelity close to 1 and actuation probability of 2/27. This scheme was found in our own design, but with analysis of its structure we succeeded to find it also in well known designs 
from \cite{Reck, Clements}. The scheme in Clements \cite{Clements} design is presented in Fig. \ref{CZClements}.  

This part of research allowed us to adjust the settings of genetic algorithms and narrow the search space, thus making a significant step towards positive result of their application.

\subsection{Experiment set 2: the search of schemes with good heralding signal}\label{3E}

The first set of experiments gave us some teasing results, but they were not new and they took too much computing time to be found. We decided to rewrite the code for scheme evaluation and search with PyTorch to employ the capabilities of modern GPUs. After we done that, the CZ gate from the previous set of experiments took 11 seconds to be found in Google Colab with GPU-enabled runtime. We assumed that 11 seconds for what previously took 11 days is a rather satisfying speedup, so we were prepared to continue this research.

For the second set of experiments we carried out the following modifications of the previous settings: 

\begin{itemize}
	\item{All schemes with $P_b$ < 1 were discarded. We were concerned about a gate with a 100\% heralding, which is very important in computation.}
	\item{We allowed phase-shifters in scheme genomes, since they are necessary for universality.}
	\item{We chose scheme parameters to match those dictated by the results of gradient descent search: the scheme depth $d=6$, the number of ancillary modes $N_a=2$, the number of ancillary photons $n_a=2$.} 
\end{itemize}

We also redefined the fitness function in order to maximize probability and discard the low fidelity schemes:

\begin{lstlisting}
def f_2(P, F):
    if F < F_min:
        return F
    return 1000 * P
\end{lstlisting}

\noindent
The threshold for fidelity this time was rather high, $F_{min} = 0.9999$, and we made sure that ancillary photons are detected in the same modes they are introduced into schemes. This significantly reduced the dimensions of the search space, while presumably eliminating its parts with low-fidelity species.

\begin{figure}[t]	
	\centering
	\includegraphics[width=3.5in]{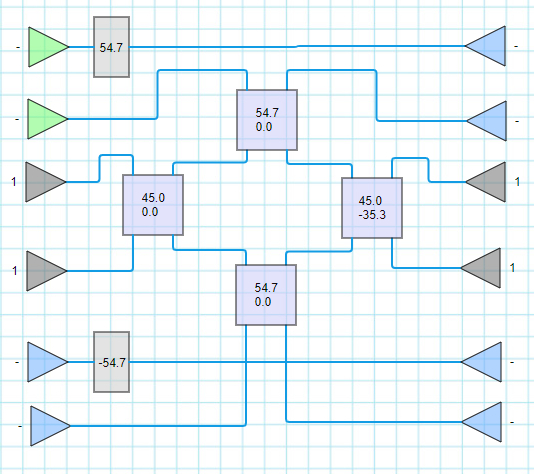}
	\caption{$CZ$, actuation probability P = $P_a$ = 2/27, $P_b$ = 1, fidelity F = 1. } 
	\label{CZ227_2}
\end{figure}

After 3 hours of search on the GPU-enabled runtime the scheme was found with fidelity $F=0.999996...$ and probability $P=0.738...$. Removing genetic garbage and attuning the angles gave us the gate (Fig. \ref{CZ227_2}) with actuation probability 2/27 and fidelity 1, which uses only 4 beam splitters and 2 phase-shifters (hereafter we will refer it CZ(2/27)). The same actuation probability and number of optical elements has the gate from \cite{Knill2002}, though the design of these schemes are different.

We provide the comparative analysis of the schemes discussed in this paper in the table below. The gate CZ(1/9) is the best in regard to actuation probability and simplicity (the number of optical elements). It has however comparatively poor heralding mechanism. As was shown in subsection \ref{2C} is not completely appropriate for computations which usually demand repetitive applications of conditional gates. The champion among the schemes with good heralding signal is gate CZ(2/27), both by the actuation probability and the scheme simplicity. 

\begin{center}
	\begin{tabular}{| c | c | c | c |}
		\hline
     	   Comparison    & CZ 1/16 &  CZ 1/9 & CZ 2/27  \\ 	
           parameters  &  Fig. \ref{figNSXCX} &  Fig. \ref{CZ93}  & fig. \ref{CZ227_2}\\
		\hline
    	Scheme depth $d$ & 8 & 3 & 4  \\ 	
		$P$ & 1/16 & 1/9 & 2/27  \\ 	
		$\bar{P_b}$ & 1 & 0.25 & 1  \\ 	
		$\bar{P_b}$ with non-PNR & & & \\ 
		 detectors & 0.689.. & 0.25 & 0.639..  \\ 
        $\bar{P_b}$ after the correction & & & \\
        procedure. & 1 & 0.25 & 1 \\
		\hline
  
	\end{tabular}
\end{center}

\subsection{Heralding mechanism correction for non-PNR detectors}\label{sec_f}

We have shown in subsection \ref{2D} that the conditional probability of gate actuation $P_b$ can decrease when non-photon-number-resolving detectors are used. In this subsection we show how one can address this problem for the gates like CZ(2/27), assuming we can distinguish only two outcomes of detection in auxiliary modes -- "no photons" or "one or more photons". 

Below we will show how to correct the situation and restore the conditional probability  $P_b$ using an additional measurement in the  proposed CZ(2/27) scheme.With analysis of all possible outcomes with presence of photons in ancillary modes one can conclude that incorrect gate actuations are associated with the leakage of photons from the qubit modes to the auxiliary ones. Such leakage is inevitably accompanied by the loss of the qubit state in the signal modes. We propose to supplement the scheme with one additional heralding measurement. After detecting the required pattern on the ancillae, one should check for coincidences at the two outputs of the control qubit modes C-C (outputs 1 and 2 counting from the top in Fig. 9). After that one should perform a similar measurement with the two outputs of the target qibit modes, T-T (outputs 5 and 6 in Fig. 9). According to the measurement results we discard the results for which there are coincidences (that is, there is lack of photons at both control/target outputs). Such a measurement restores the value of the conditional probability of the correct scheme actuation $P_b = 1$.

It should be noted that such a check is possible due to the peculiarity of our scheme. With correct heralding signal on ancillae one can never detect photons in signal mode, if no photons were introduced into this signal mode. The photon from the control mode cannot get into the target one and vice versa. This fact indicates another remarkable feature of this check: it does not interfere with the scalability of calculations, that is, it can be carried out once at the end of calculations, after a chain of several gates.

Let's consider the described procedure in more details. For example, let the state $\ket{0101}$ be initiated on the input of the CZ(2/27)  gate. At the output of the scheme, we check the pattern on the ancillae. Among the possible erroneous outcomes that we could not distinguish, there is an outcome when two photons are present into one of the ancillae, and another photon --  into the other. At the same time, in one of the signal modes of the output state (for example, in the target) there will be 0 photons in both channels. Without performing any additional actions, we send the result to the input of the second similar scheme CZ(2/27). State $\ket{0100}$ evolves in the second scheme, after which we check the pattern again on the ancillae of the second scheme. The chain of gates can be of the required length. For the selected variants, we check not only the heralding pattern on the ancillae, but also the matches in the C-C and T-T channels, at the output of the chain. All options will be eliminated based on the presence of a match "no photons"-"no photons". That is, the presence of an incorrect operation in one of the gates could always be distinguished at the output. 

It is important to note that although the proposed check looks similar to the one implemented in the CX(1/9) and CZ(1/9) schemes, these schemes do not allow us to filter out incorrect actuations at one gate from the chain by the final measurement. For example, taking the state $\ket{1010}$ at the input of the CZ(1/9) gate as one of the possible options, we will get the state $\ket{2000}$ at the output. According to the ancillae pattern for this scheme, we do not recognize an incorrect actuation. If such a state income to the second similar gate from the chain, it can turn into $\ket{1010}$ at the output, and by monitoring coincidences we will not recognize that one of the gates in the chain worked incorrectly. Such a defect is associated with the possibility of photon migration from the control mode to the target mode and vice versa during the evolution of the state.
 Note that the CZ(1/16) protocol and CZ(2/27) gates does not allow photon migration between the target and control modes, which means that even if there are non photon number resolving detectors, the heralding mechanism in this scheme can also be adjusted by an additional measurement of coincidences in the signal modes at the end of the calculation without compromising scalability.

\section{Conclusion}

In this research we presented the results of genetic algorithms application to the problem of heralded conditional quantum gate linear optic schemes search. With this approach we managed to find two high fidelity optical schemes. One of these schemes has the best currently known actuation probability $P=1/9$ with no regard to the  heralding signal quality, and the other has the best actuation probability $P=2/27$ among the schemes with good ($P_b=1$) heralding mechanism.   

The result of the first set of experiments convinced us that 1/9 is the best actuation probability for the conditional gates in KLM-Protocol, assuming we don't have entangled ancillary photons. Increasing the number of ancillary as well as the number of photons in the ancillary modes only worsens the situation. The gate CZ(1/9) found with an extensive search has the same characteristics, as the analytically deduced in \cite{CNOT2003} gate CX. 

We also note that the gate actuation probability is not the only useful characteristic of an optical scheme. The quality of correct actuation heralding event becomes much more important when one needs to perform computations placing several conditional gates in line with no intermediary measurements (except those in ancillary modes). We introduced the simple metrics which allows to compare the schemes with less than 100\% heralding reliability, regardless of the logical basis and applied this metrics for comparison of the discussed optical schemes. 

In the second set of experiments we placed the restriction on the schemes to have good heralding mechanism. As a result we found the gate CZ(2/27) which has $P_b=1$, fidelity $F=1$ and actuation probability $P=2/27$, the same characteristics as in \cite{Knill2002}, though with different (and slightly simpler) design.  

The results obtained show the prospects of the developed method of scheme analysis – both circuits with record characteristics for certain parameters were found as a result of the genetic search.

All of the considered gate schemes require PNR detectors for obtaining the correct heralding pattern, which are questionably available resource at the present time. We analysed the behaviour of the considered optical schemes under the assumption that we can't distinguish the actual number of detected photons. The heralding signal quality decreases in this case, thus we introduced the procedure to restore it for the gate CZ(2/27) and that found by Knill (\cite{Knill2002}). This procedure allows scaling of computations and requires the signal photons detection only at the end of computations where such detection is natural.  

The fact that the schemes with best known characteristics were found with no additional insight shows that genetic algorithms are the perspective and useful instrument for linear optical schemes design for quantum computing. We believe that it will be successfully used further for this type of tasks and produce new fruitful results. 

The code of all experiments is available on GitHub \cite{galopy}. 

\section{Acknowledgements}
This work was financially supported by the Russian Science Foundation (grant No. 22-22-00022).

\end{document}